\RequirePackage{snapshot}

\documentclass[utf8]{FrontiersinHarvard} % for articles in journals using the Harvard Referencing Style (Author-Date), for Frontiers Reference Styles by Journal: https://zendesk.frontiersin.org/hc/en-us/articles/360017860337-Frontiers-Reference-Styles-by-Journal
\usepackage{url,hyperref,lineno,microtype,subcaption}
\usepackage[onehalfspacing]{setspace}
\usepackage{comment}
\usepackage{multirow}
\usepackage{gensymb}
\usepackage{wrapfig}
\usepackage{float}
\usepackage[ruled,vlined]{algorithm2e} 
\usepackage{siunitx}

\def\keyFont{\fontsize{8}{11}\helveticabold }
\def\firstAuthorLast{Shemonti {et~al.}} %use et al only if is more than 1 author

\def\Authors{Abida Sanjana Shemonti\,$^{1}$, Joshua D. Eisenberg\,$^{3}$, Robert O. Heuckeroth\,$^{3,4}$, Marthe J. Howard\,$^{5}$, Alex Pothen\,$^{1}$ and Bartek Rajwa\,$^{2*}$}

\def\Address{$^{1}$ Purdue University, Department of Computer Science, West Lafayette, IN, USA\\
$^{2}$ Purdue University, Bindley Bioscience Center, West Lafayette, IN, USA\\
$^{3}$ The Children's Hospital of Philadelphia, Philadelphia, PA, USA\\
$^{4}$ University of Pennsylvania, Perelman School of Medicine, Philadelphia, PA, USA\\
$^{5}$ University of Toledo, Department of Neurosciences, College of Medicine and Life Sciences, Toledo, OH, USA
}

\def\corrAuthor{Bartek Rajwa}\def\corrEmail{brajwa@purdue.edu}

\begin{document}
\onecolumn
\firstpage{1}

\title[Generative Modeling of the Enteric Nervous System]{Generative Modeling of the Enteric Nervous System Employing Point Pattern Analysis and Graph Construction} 

\author[\firstAuthorLast ]{\Authors} %This field will be automatically populated
\address{} %This field will be automatically populated
\correspondance{} %This field will be automatically populated

\extraAuth{}% If there are more than 1 corresponding author, comment this line and uncomment the next one.
%\extraAuth{corresponding Author2 \\ Laboratory X2, Institute X2, Department X2, Organization X2, Street X2, City X2 , State XX2 (only USA, Canada and Australia), Zip Code2, X2 Country X2, email2@uni2.edu}

\maketitle

\begin{abstract}
We describe a generative network model of the architecture of the enteric nervous system (ENS) in the colon employing data from images of human and mouse tissue samples obtained through confocal microscopy. Our models combine spatial point pattern analysis with graph generation to characterize the spatial and topological properties of the ganglia (clusters of neurons and glial cells), the inter-ganglionic connections, and the neuronal organization within the ganglia. We employ a hybrid hardcore-Strauss process for spatial patterns and a planar random graph generation for constructing the spatially embedded network. We show that our generative model may be helpful in both basic and translational studies, and it is sufficiently expressive to model the ENS architecture of individuals who vary in age and health status. Increased understanding of the ENS connectome will enable the use of neuromodulation strategies in treatment and clarify anatomic diagnostic criteria for people with bowel motility disorders.

\tiny
 \keyFont{ \section{Keywords: Enteric Nervous System, Connectome, Colon, Spatial Point Process, Spatially Embedded Random Network} } %All article types: you may provide up to 8 keywords; at least 5 are mandatory.
\end{abstract}

\newpage

\section{Introduction}   

\subsection{Architecture and the connectome of the enteric nervous system}

\begin{wrapfigure}[24]{r}{0.5\textwidth} 
	\centering
	\includegraphics[width=0.5\textwidth]{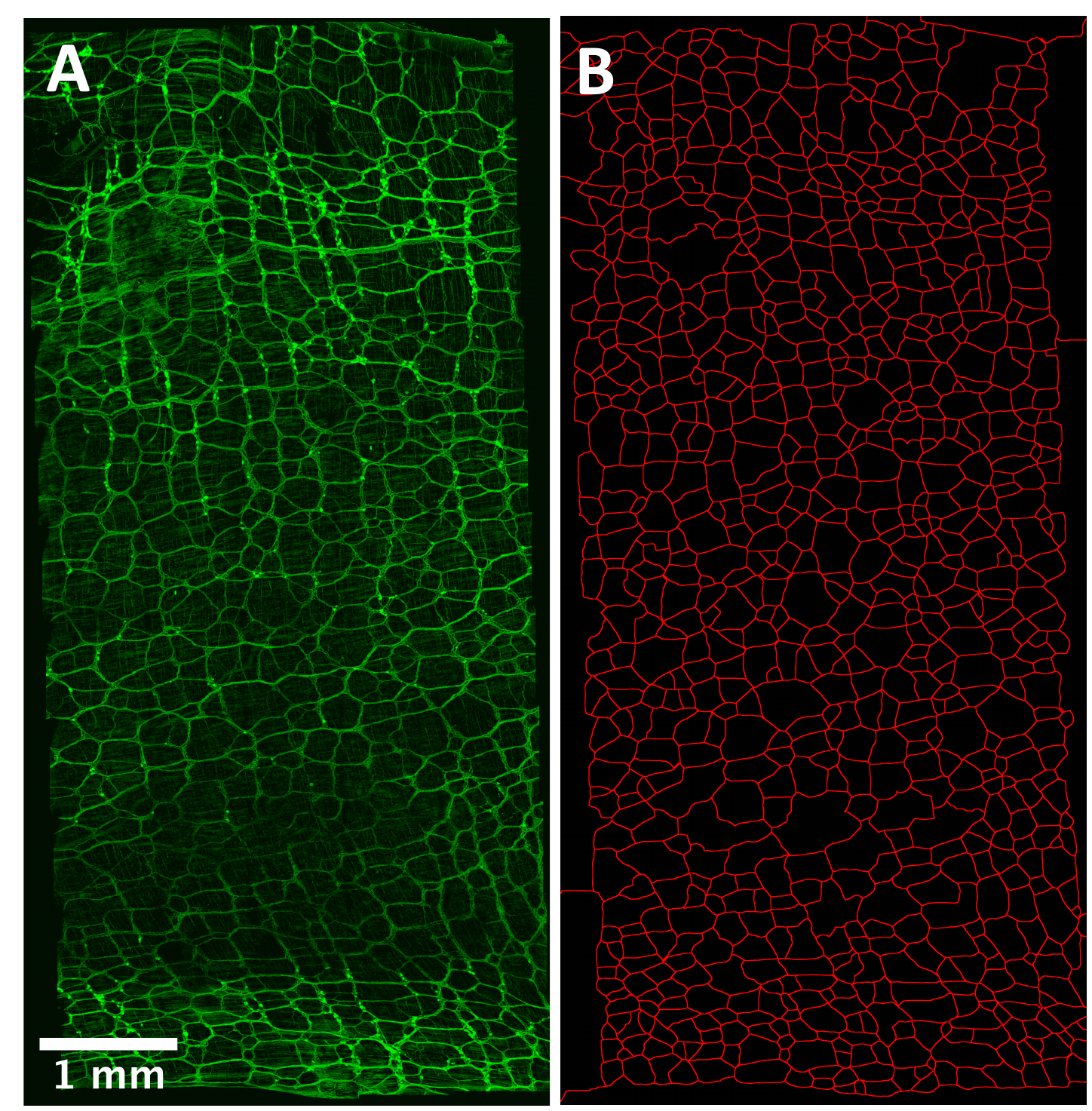}
	\caption{(A) A mouse colon myenteric network; only the inter-ganglionic connections are shown (green). (B) Visualization of a spatial network created via a multistep image processing that involves FFT filtering/denoising, morphology-based segmentation, and skeletonization.}
	\label{fig:mouse_ENS}
\end{wrapfigure}

We describe a generative network model of the enteric nervous system (ENS) in the colon, developed with data obtained from confocal images of human and mouse tissue samples. Our models combine spatial point pattern analysis (SPP)  with graph generation to characterize the spatial and topological properties of the ganglia (clusters of neurons and glial cells), the inter-ganglionic connections, and the neuronal organization within the ganglia. We show that our approach, initially informed by the mouse colon ENS neuroanatomy (Figure~\ref{fig:mouse_ENS}), can simulate a variety of human samples differing in age and anatomical pathologies \citep{Graham2020, Nestor-Kalinoski2022}. We also show that the combination of  SPP and graph generation approaches suffices to produce models exhibiting crucial properties of ENS anatomy, morphology, and circuitry.

Despite the ongoing focus on mapping the central nervous system (CNS) connectome \citep{Bassett2017, Vecchio2017, Swanson2010}, the peripheral nervous system (PNS) architecture in general, and enteric nervous system (ENS) in particular, have been largely disregarded. Besides the studies performed using simple biological models \citep{Bezares-Calderon2016}, not much effort has been directed towards describing, quantifying, modeling, and understanding the topology and architecture of the PNS~\citep{irimia_mapping_2021}. While the ENS has been described in terms of neuron and glial density or other fundamental characteristics, anatomic descriptors of ENS networks have not been well defined, so it is difficult to recognize when these ENS networks are disorganized \citep{Graham2020}.

Defining ENS network properties is valuable because diverse genetic, toxic, infectious, metabolic, nutritional, and inflammatory factors can disrupt ENS function to cause human diseases collectively called ``bowel motility disorders''~\citep{schneider2019unexpected, furness2016integrated}. When the ENS is abnormal, bowel dysfunction can be debilitating or deadly since the ENS controls most aspects of bowel function in response to local stimuli. To monitor and regulate bowel function, the ENS is equipped with approximately as many nerve cells as the spinal cord, at least fourteen different neuron types, and nearly four times as many glial cells~\citep{schneider2019unexpected, Graham2020, wright2021scrna}. These ENS neurons and glia are distributed along the bowel in interconnected clusters called ganglia, embedded in two ganglionated plexi (intersecting networks): the myenteric plexus and the submucosal plexus. The primary function of the former is the control of muscle contraction and relaxation, while the role of the latter is to regulate epithelial function, blood flow, and the bowel immune system. ENS sensory neurons respond to luminal nutrients and stretch~\citep{neunlist2014nutrient}. The ENS influences almost every bowel cell type and ENS cells respond to signals from many other cells~\citep{schneider2019unexpected}. These include close interactions of ENS with smooth muscle, interstitial cells of Cajal, PDGFR$\alpha$ cells, enteroendocrine cells, blood vessels, muscularis macrophages, and other bowel immune system cells.

The most dangerous ENS diseases are Hirschsprung's disease (HSCR)~\citep{heuckeroth2018hirschsprung, gosain2015hirschsprung, langer2011hirschsprung}, where ENS is absent from distal bowel and chronic intestinal pseudo-obstruction (CIPO)~\citep{di2017chronic}, where ENS is present but defective. HSCR and CIPO cause profound bowel dysmotility that can be life-threatening, requiring surgery or intravenous nutrition starting in infancy. Bowel dysmotility also occurs in Parkinson's disease, gastroparesis, achalasia, inflammatory bowel disease, and irritable bowel syndrome. These diseases are challenging to treat, and mechanisms are incompletely understood, limiting targeted therapy. A clear understanding of how these human diseases impact ENS anatomy or how anatomy might predict function would be valuable as we design new diagnostic and therapeutic strategies for people with bowel motility disorders.

Thus, our long-term goal is to describe and analyze the ENS connectome structure using graph-theoretic, network-based, and spatial-point pattern (SPP) process-related analytic approaches to develop robust mathematical models that distinguish normal from abnormal ENS anatomy. This is important because there is currently no definition of what morphology and connectivity patterning constitute normal or disorganized ENS networks. The work on the CNS connectome has demonstrated the value of the graph-analytic perspective for network quantification. Although graph-analytic approaches have been used in biology to model relationships between abstract entities (organisms, individuals, metabolites, regulatory systems, etc.),  with a few simple exceptions, graph analysis has rarely been employed to characterize the spatial organization of anatomical structures. On the other hand,  SPP techniques have been occasionally used in neuroscience since the seminal work of Ripley (\citeyear{Ripley1977}). For instance, ~\cite{Diggle1994} utilized spatial analysis to investigate the distribution of pyramidal neurons in the cingulate cortex of normal subjects and schizophrenic patients. 

\subsection{Spatial point patterns and spatial networks} 

We briefly introduce the tools utilized in the process of building our model. The two critical conceptual components are the notions of \emph{spatial point patterns} and \emph{spatial graphs} or \emph{networks}.

A \emph{spatial point pattern} (SPP) $\textbf{X}$ is a collection of spatial locations assigned to objects or events of interest in 2D or 3D space~\citep{baddeley2015spatial, moller2003statistical, Stoyan2006, jafari2010spatial}. The points in an SPP (i) may be of different types (multitype point pattern),  (ii) could be linked to auxiliary information or characteristics (marked point pattern), or (iii) could be linked to the space of interest (covariates). SPP analysis studies the points' spatial arrangement and aims to identify trends that characterize the patterns. It is common practice to compute summary statistics like nearest-neighbor distance, empty-space distance, or pair correlation to examine point patterns. A key component of an SPP is the underlying method or model for generating points. Every SPP can be thought of as an outcome of a spatial point process, a generative statistical model with predefined parameters describing a trend's formation. Two crucial parameters for a point process are \emph{intensity} and \emph{interaction}. The intensity ($\lambda$) of a point pattern is the average number of points per unit area, and it could be constant across space (stationary point pattern), or it might vary according to an intensity function (non-stationary point pattern). The interaction parameter describes the influence that the points have on each other. If the points are independent, the outcome is referred to as complete spatial randomness (CSR) and could be modeled with the Poisson point process. If the points exhibit positive interaction (spatial attraction), they can be modeled with various cluster and Cox processes. If they display negative interaction (spatial inhibition), they may be modeled with Gibbs processes~\citep{baddeley2015spatial}. 

A \emph{spatial network} is a graph $G(V, E)$ where the set of vertices $V$ may be conceptualized as a spatial point pattern, and the set of edges $E$ represents some context-dependent interaction between the points, constrained by geometry and linked to a distance~\citep{barthelemy2018morphogenesis, huang2014navigation,guimera2005worldwide, marshall2018street}. The class of random networks developed by considering the underlying spatial constraints is sometimes referred to as spatially embedded random networks (SERNs)~\citep{hackl2017generation, barnett2007spatially, parsonage2017fast}. Spatial networks often share similar traits such as scale-free nature~\citep{albert2002statistical} and small-world characteristics~\citep{watts1998collective}. Typically in these networks, closely located vertices are more likely to be connected than more spatially distant vertices. Vertices also tend to have low degrees. A specific class of spatial networks is a \emph{spatial planar network} that can be drawn in the plane so that its edges do not intersect \citep{Barthelemy2011}.

A number of measures have been defined to describe  spatial networks. A key measure is the \emph{clustering coefficient}, which  for a vertex $i$ of degree $k_i$ is defined as:
\begin{equation}\label{eq:cc}
	C(i) = \frac{E_i}{k_i(k_i-1)/2},
\end{equation}
where $E_i$ is the number of edges among the neighbors of vertex $i$. The average value of the clustering coefficients of all the vertices is the clustering coefficient of the network. This metric is highly relevant for spatial networks with planarity, as this property increases the probability that closer vertices will be connected. Planarity results in a higher clustering coefficient compared to the values in a  random Erd\"{o}s–R\'{e}nyi graph. Other relevant measures adopted from geographical networks are $\alpha$ (meshedness)~\citep{buhl2006topological}, $g$ (density of network)~\citep{kansky1963structure} and $\psi$ (compactness)~\citep{courtat2011mathematics}, which for planar networks are defined as:
\begin{equation}
	\alpha=\frac{E-N+1}{2N-5}, \text{ }g=\frac{E}{3N-6}, \text{ }\psi=\frac{4A}{(\ell_T-2\sqrt{A})^2},
\end{equation}
where $N$ is the number of vertices, $E$ is the number of edges, $A$ is the total area occupied by the spatially embedded network, and $\ell_T$ is the total length of all the edges. Note that a planar graph can have at most $3N-6$ edges, which is the denominator of $g$; the numerator and denominator of $\alpha$ have the value $(N-1)$ subtracted from them (it is the number of edges in a tree with $N$ vertices). 
All three measures have values in the interval [0, 1]. The value of meshedness ($\alpha$) is 0 for trees and 1 for maximal planar graphs (planar graphs with the maximum number of edges). The density of the network ($g$) is the ratio of the number of edges present in the graph to the maximum number of edges that could be in the planar graph. Compactness ($\psi$) represents how much area is filled with edges~\citep{barthelemy2018morphogenesis}. 

\begin{figure}
    \centering
    \includegraphics[width=\textwidth]{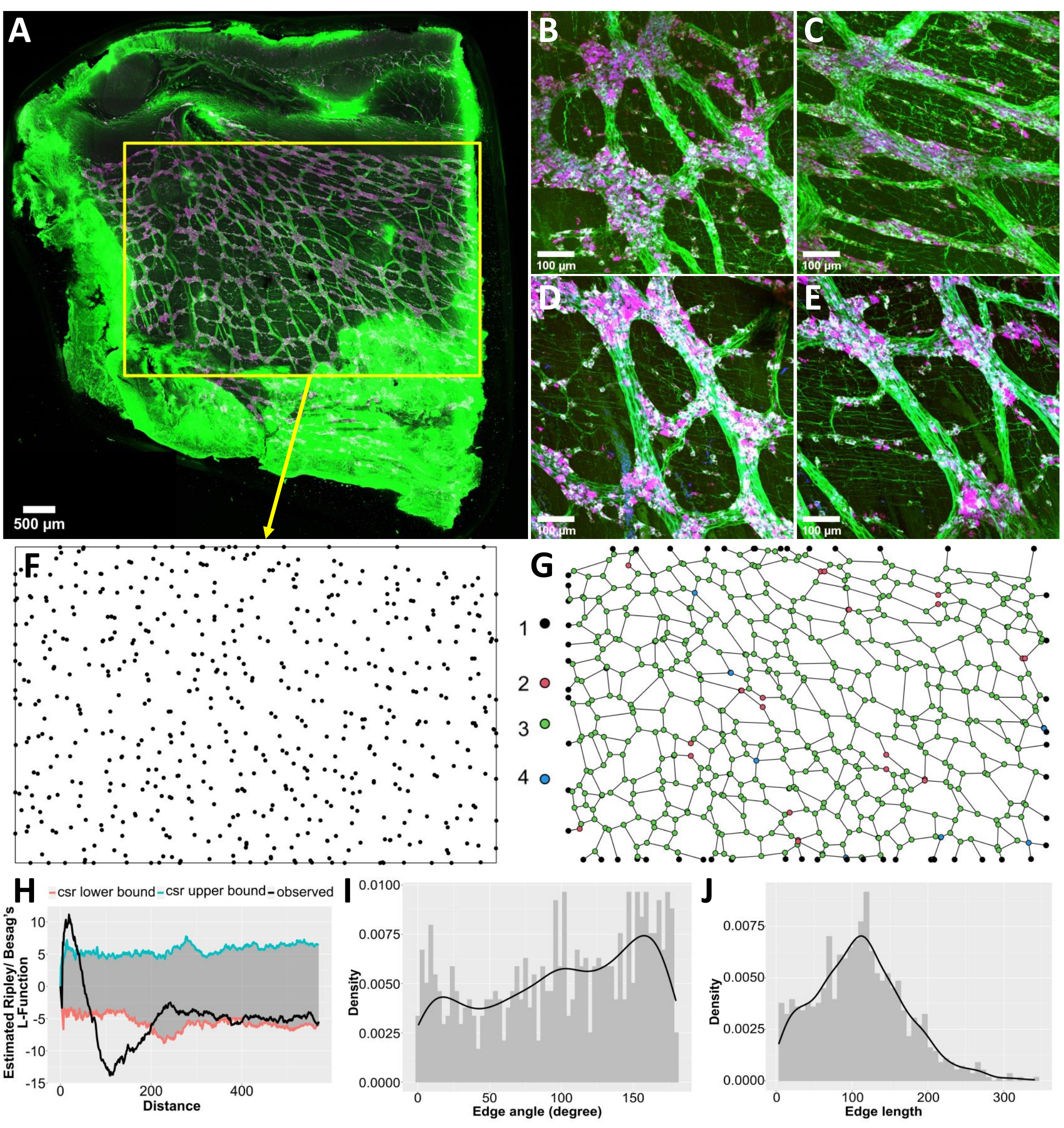}
    \caption{Characterization of the enteric nervous system. (A) The flattened multichannel z-stacks of the myenteric plexus of ENS sample 2475-P, labeled with HuC/D (purple) and nNOS (green). (B-E) Four magnified regions taken from different quadrants of (A) displaying enteric neurons cell bodies (purple) within the ganglia and the inter-ganglionic connections (green). (F) The spatial point pattern of the ganglia centers of the region of interest (outlined in yellow) in (A). (G) The corresponding spatial network with the vertices colored according to their degrees (labels on the left). (H) The L-function plot characterizing the spatial arrangement of the ganglia centers. The shaded area between the upper (teal) and lower (orange) bounds represents the range of values expected from complete spatial randomness; hence this plot indicates a prominent level of non-random spatial organization in the observed point pattern (black) up to a certain distance (pixels). (I, J) Distribution of the edge angle (degree) and the edge length (pixels) in the spatial network, respectively. For all distances and lengths related to (A) unit pixel = 1.5743 \textmu m and (B-E) unit pixel = 0.7872 \textmu m.} 
    \label{fig:human_ens}
\end{figure} 
\section{Methodology}
 
In the simplest conceptual approximation of the ENS network, its structure resembles a planar spatial network with the positions of the ganglia generated from a spatial point process with negative interactions (see Figure~\ref{fig:mouse_ENS}). The ganglia centroids constitute the vertices of the network, and interganglionic connections are the network edges. Macroscopically, in terms of geometric embedding, the ENS consists of two major interconnected spatial networks (myenteric and submucosal plexi) embedded into concentric cylinders, with the second network internally comprised of highly linked subnetworks \citep{nagy2017enteric, furness2012enteric}. However, for the purpose of the modeling, we focus on the myenteric plexus only.

\subsection{Biological data}  

For the development of our initial models, we used confocal images of mouse and human ENS in which the tissue was immunolabeled, allowing for easy visualization of different anatomical structures (see Figure~\ref{fig:mouse_ENS}A and Figure~\ref{fig:human_ens}A). The interganglionic network in the mouse middle colon resembles the architecture of connections observed in humans.

Figure~\ref{fig:human_ens}A shows the flattened optical z-stacks (combination of multiple images taken at different focal distances) of the myenteric plexus collected from the proximal margin of a region of left colon removed from a five-month-old child with Hirschsprung disease (sample ID: 2475-P). The ENS structures were labeled using antibodies to the human neuron protein HuC/D (RRID: AB\_221448) and to neuronal nitric oxide synthase nNOS (RRID: AB\_91824) and then visualized using secondary antibodies Alexa Fluor 594 (purple enteric neuron cell bodies, RRID: AB\_141633) and Alexa Fluor 488 (green inter-ganglionic connections, RRID: AB\_253579). Higher magnification confocal images are in Figures~\ref{fig:human_ens}B-E. The details of the methods used to capture the images of human ENS samples are described in previously published reports \citep{Graham2020}. The flattened z-stacks of the myenteric plexus were further processed using ImageJ (Fiji) software for manual segmentation and semi-automated extraction of the location of the ganglia centers and the connections of the networks \citep{schindelin2012fiji}. The semi-automated procedure involves denoising, Fourier filtering, and morphology-based segmentation. Subsequently, the pre-preprocessed network images were processed with Fiji Skeleton Analyzer, which provided the length of the interganglionic connections (network edges), location of the edges, distances between endpoints, and other helpful network features.

\subsection{Characterization of the ENS network} 

\subsubsection{ENS ganglia} 

We use measures commonly employed in the exploratory analysis of spatial point patterns~\citep{baddeley2015spatial, moller2003statistical, Stoyan2006, jafari2010spatial} to characterize the location of ENS ganglia. A useful summary function, the Ripley K-function, in Besag's L-function format (shown in Equation~\ref{eq:l-funct}) allows us to demonstrate the non-random spatial arrangement of ganglia by computing the cumulative distribution of the neighbors of the observed data points and comparing it with known alternative arrangements, such as complete
spatial randomness (CSR).

\begin{equation}\label{eq:l-funct}
\begin{gathered}
  K(r) = \frac{W}{{n(n - 1)}}\sum\limits_{i = 1}^n {\sum\limits_{j = 1,j \ne i}^n {\nu ({d_{ij}} \leqslant r){e_{ij}}(r)} } , \hfill \\
  L(r) = \sqrt {\frac{{K(r)}}{\pi }}  - r,\; \hfill \\ 
\end{gathered}
\end{equation} 
where $\nu(.)$ is an indicator function that equals $1$ if the argument is true, and otherwise is $0$. Here 
$n$ is the number of points;  $W$ is the observation window;  $e_{ij}$ denotes weights for edge correction;  and $r$ is the distance. Positive values  of the L-function depict spatial attraction, and negative values describe spatial inhibition. 

Figure~\ref{fig:human_ens}H shows the L-function plot for the spatial point pattern of ganglia centers of the ENS network shown in Figure~\ref{fig:human_ens}A, where the black curve falls visibly outside of CSR bounds demonstrating that the arrangement of ENS ganglia is non-random. The curve also displays significant spatial interaction (represented by one upward and one significant downward peak) up to a certain distance. Hence the organization of ENS ganglia shows a hybrid form of spatial inhibition occurring up to a certain distance. This behavior suggests that the arrangement of the ganglia might be modeled with Gibbs processes like a hybrid hardcore-Strauss defined by the intensity, interaction parameter, interaction distance, and hardcore distance \citep{baddeley2015spatial}.

\subsubsection{ENS network} 

The spatial graph in Figure~\ref{fig:human_ens}G represents the ENS network in Figure~\ref{fig:human_ens}A with the ganglia centers as vertices and corresponding inter-ganglionic connections as edges. The network is planar, and the vertices have a low degree (with a mode of 3), resembling a small-world model. The network has low meshedness ($\alpha$=0.19), moderate network density ($g$=0.46), and high compactness ($\psi$=0.995). The clustering coefficient of the network is 0.0097. The degree-one vertices are located at the sample boundary because of the dissection artifact. We also characterize the network in terms of the angle and the length of the edges. The distribution of the angle of the edges (see Figure~\ref{fig:human_ens}I) is multimodal, with modes approximately at 10, 96, and 169 degrees, which gives the entire network a certain dominant orientation. Figure~\ref{fig:human_ens}J shows the distribution of the length of the edges with an average length of 109 pixels (unit pixel = 1.5743 \textmu m). 

\subsection{Generative model of the ENS network} 

The characteristics described above provide a mathematical description of the ENS network. Now we develop a well-parameterized generative model that provides realizations closely matching those observed characteristics. Our generative model incorporates a three-step process: (i) generation of starting positions for the ganglia (which we refer to as ``ganglia centers'' although they may not technically be geometrical centers of the objects), (ii) generation of the edges (inter-ganglionic connections), and (iii) generation of the neurons' centroids (within the body of the ganglia). It is important to emphasize that the sequence of generative steps in the model does not necessarily follow the biological sequence of events during ENS development and maturation. It was chosen for algorithmic convenience. Specifically, we envision a generative model in which the ganglia are created via some pairwise point interaction process~\citep{strauss1975model}. We simulate ganglia centers with the hybrid hardcore-Strauss process (HSP) with intensity parameter $\beta$, interaction parameter $\gamma$, interaction radius $R$, and hardcore distance $H$. In this process, each point contributes a factor $\beta$ to the probability density of the point pattern, and each pair of points closer than $R$ units contributes a factor $\gamma$ to the density. Let $x$ be a point pattern and $t(u,R,x)$ be the number of points in $x$ which lie within distance $R$ of the location $u$. The conditional intensity is defined as: 
\begin{equation}\label{eq:cond_intensity}
\lambda(u|x) = 
    \begin{cases}
      0 & \text{if $||u-x_i|| < H, \text{for some} \ i$},\\
    \beta \gamma^{t(u,R,x)} & \text{else \ if } H\leq||u-x_i||\leq R, \text{for some } \ i,\\
    %0 & \text{ if $d< H$}\\
    \beta & \text{else\ if $||u-x_i|| > R, \text{for\ all}\  i$}. 
    \end{cases} 
\end{equation}
The HSP point pattern is stationary when the intensity parameter ($\beta$) is a positive numerical constant. The intensity parameter ($\beta$) can be a function of certain features of the points (i.e., $x$-coordinates, $y$-coordinates) resulting in a  non-stationary HSP point pattern. Because ganglia occupy  physical space, the incorporated hardcore distance disallows overlap and unrealistically close neighborhoods between structures. Although close pairs of points are allowed above the distance $H$, they can still be penalized in the model.

For the edge generation process, we choose a combination of a deterministic and a random connection model following the methodology by~\cite{hackl2017generation}. Let us denote by $\textbf{X}$ an instance of the HSP in $\mathbb{R}^2$, by $C_d(\textbf{X})$ a deterministic connection model, and by $C_r(\textbf{X}, c)$ a random connection model with connection function $c(.)$. We propose the deterministic connection model to be a Delaunay triangulation, which is the dual of Voronoi regions and is based on minimum spanning trees~\citep{hackl2017generation, marchette2005random}. We construct the initial template for the inter-ganglionic connections $G(V, E)$, where $V=\textbf{X}$ and $E=C_d(\textbf{X})$ is the Delaunay triangulation on $\textbf{X}$. This step returns a  maximal planar spatial graph (planar graph with most number of edges). Subsequently, we rewire (following the terminology of~\cite{watts1998collective}) every edge $(v_i, v_j)\in E$ according to the random connection model $C_r(\textbf{X}, c)$. We pick $C_r(\textbf{X}, c)$ to be a rejection sampling model so that the vertices are connected with a certain probability~\citep{parsonage2017fast}. Let $G_0(V_0, E_0)$ be a real ENS network, and $D_f(.)$ the probability distribution of a set of one or more features $f$ of a given network, such as
edge length, edge orientation, the degree of vertices, etc. The purpose of this distribution is to compute the probability of the edge $(v_i, v_j)$ sampled from the edges of the Delaunay triangulation given the probability distribution of the real ENS network for the selected $f$. We denote $f_{ij}$ as the value of the selected feature set $f$ for the edge $(v_i, v_j)$. The random connection model $C_r(\textbf{X}, c)$ keeps the edge $(v_i, v_j)$ in $E$ for which the connection function $c(v_i, v_j)$=1 and removes the other edges to form the final realization $G(V, E')$. We formulate the connection function as follows:
\begin{equation}
    c(v_i, v_j)= 
    \begin{cases}
        1 & \text{if } \frac{P(f_{ij}|D_f(G))}{P(f_{ij}|D_f(G_0))} \leq 1,\\
        0 & \text{otherwise.}
    \end{cases}
\end{equation}

\begin{algorithm}[t]
    \SetAlgoLined
    \KwIn{HSP point pattern $\textbf{X}$}
    \KwOut{Edge set $E'$}
     $E' \leftarrow \emptyset$ \;
     $E \leftarrow \text{Delaunay Triangulation }(\textbf{X})$ \; 
     \For{each edge $(v_i, v_j)\in E$}{
        assign weight $w_{ij}$ using Equation~\ref{eq:edge_weight} \;
     }
     \While{\text{no more edges can be rejected}}{
      pick an edge $(v_i, v_j)\in E$ with probability $w_{ij}$s \;
      calculate $p_{ij} \leftarrow P(f_{ij}|D_f(G_0))$\;
      calculate $r \leftarrow P(f_{ij}|D_f(G))$ \;
      \eIf{$r \le p_{ij}$ or $getsDisconnected(E\setminus(v_i,v_j))$}{
        keep the edge $(v_i, v_j)$\;
       }{
        $E \leftarrow E \setminus (v_i, v_j)$ \;
        \For{each edge $(v_i, v_j)\in E$}{
            update weight $w_{ij}$ using Equation~\ref{eq:edge_weight} \;
        }
       }
     }
     $E' \leftarrow E$ \;
    \caption{Generalized Spatial Network Generation}
    \label{algo:1}
\end{algorithm} 

A generalized outline of the edge generation process is shown in Algorithm~\ref{algo:1}. After constructing the Delaunay triangulation, we assign a $0$-$1$ normalized weight to each edge as follows:
\begin{equation}\label{eq:edge_weight} 
\begin{gathered}
  {w}_{ij}' = k_{v_i} + k_{v_j}, \\
  w_{ij} = \frac{w_{ij}'- min(w')}{max(w')-min(w')},
\end{gathered}
\end{equation} 
where $w_{ij}'$ is the unnormalized weight of the edge $(v_i, v_j)$; $k_v$ is the degree of the vertex $v$; $w'$ is the set of all unnormalized edge weights; $w_{ij}$ is the $0$-$1$ normalized weight of the edge $(v_i, v_j)$; and $min(.)$ and $max(.)$ are the functions returning the minimum value and the maximum value of a given set, respectively. The 0-1 normalized edge weights work during the sampling process as a probability to pick an edge whose end vertices (either or both) have high degrees. For the stopping condition of the sampling process (mentioned as \emph{no more edges can be rejected} in Algorithm~\ref{algo:1}) we incorporate two criteria: (i) we stop if any one of the three network measures, meshedness, network density or compactness, for the spatial network under development become lower than the measures of the real ENS network, and (ii) we stop if no edges can be rejected for a significantly large number of iterations. To address the dissection artifact, we only allow vertices at a certain distance ($\epsilon$) from the boundary of the point pattern to have a degree of one or two. To compare the spatial networks resulting from Algorithm~\ref{algo:1} for different choices of the feature set, we use the earth mover's distance (EMD). The EMD, first introduced for image analysis by Rubner et al.~(\citeyear{rubner2000earth, rubner1998metric}), has an equivalent metric on probability distributions known as Mallows, or Kantorovich-Wasserstein distance~\citep{mallows1972note, levina2001earth}.    
 
Although ganglia centers can play the role of dimensionless vertices in the simplest model, we extend the approach further by simulating the neurons present in the ganglia via another use of spatial point processes. Therefore, in the simulation's last step, our model produces the locations of neurons' centers in the ganglia. Figures~\ref{fig:human_ens}B-E give close views of the arrangement of the neurons (cell bodies) within the ganglia of a real ENS. The ENS network structure appears to correspond to the neurons' spatial distribution. Therefore, we generate the points simulating the neurons' centers using an inhomogeneous HSP process. The edges of the simulated inter-ganglionic network serve as a basis for an underlying intensity profile, which introduces the inhomogeneity. The objective is to create intensity profiles representing higher densities near ganglia locations, gradually decreasing along the edges of the network. To achieve this, we generate a 2D binary image (pixel matrix) of the network $G(V, E')$, and then execute a series of morphological image transformations (several dilations followed by Euclidean distance mapping using Fiji to construct a greyscale mask used as the probability profile for the inhomogeneous HSP~\citep{leymarie1992fast}. This image transformation produces an intensity profile that influences the spatial extent of the ganglia. With the profile generated, the inhomogeneity of the process is achieved by applying multiple random thinning operations using the generated thinning surface~\citep{baddeley2000non, jensen2001review}.  
 
\section{Experimental results and discussions} 

\subsection{Simulation of the generative model}

\begin{figure}
    \centering
    \includegraphics[width=\textwidth]{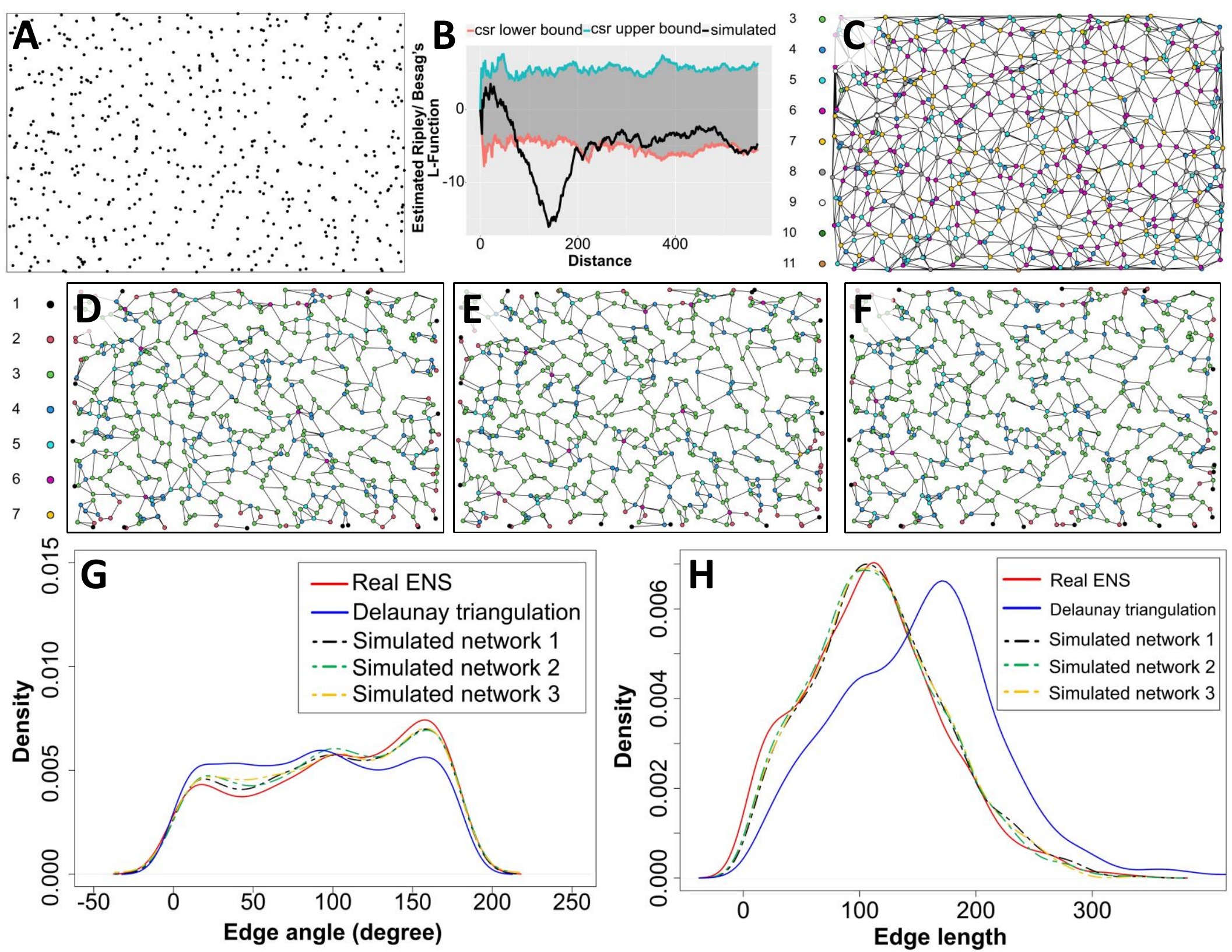}
    \caption{Experimental results for simulated realizations of the ENS. (A) The ganglia centers simulated from the non-stationary hybrid HSP model. (B) The L-function plot characterizing the spatial arrangement of the simulated ganglia. The shaded area between upper (teal) and lower (orange) bounds represents the range of values expected from complete spatial randomness, indicating a significant level of non-random spatial organization in the simulated point pattern (black) up to a certain distance (pixels). The trend in the L-function is similar to the real ENS. (C) The Delaunay triangulation constructed with the simulated ganglia centers as the vertices; the vertices are colored according to their degrees (labels on the left). (D-F) Three different realizations of simulated ENS network for the same underlying ganglia centers; the vertices are colored according to their degrees (labels on the left). (G, H) Distribution of the edge angle (degree) and edge length (pixels) of the real ENS network, intermediate triangulation, and the three simulated networks. For the distances and lengths in this experimental results unit pixel = 1.5743 \textmu m.} 
    \label{fig:simulated_ens}
\end{figure}

In our work, we use existing comprehensive tools for spatial point process statistics and graph modeling, including R libraries \textit{spatstat}~\citep{baddeley2015spatial, baddeley2020spatial} and \textit{igraph} \citep{Csardi2006}. We fit the hybrid hardcore-Strauss process (HSP) to the spatial point patterns formed by the centers of the ENS ganglia in such a way that the interaction parameter confirms spatial inhibition ($\gamma\leq$ 1). We use the function \textit{spatstat::ppm} which includes a computationally efficient technique, based on logistic regression, for fitting Gibbs point process models to SPP data~\citep{baddeley2014logistic, Diggle1994, jensen1994asymptotic}. Although we experimented with both stationary and non-stationary HSP models, we report the non-stationary HSP results only as they are more appropriate for the generation of ENS ganglia centers. Equation~\ref{eq:beta} shows the intensity function of the non-stationary HSP model.
\begin{equation}\label{eq:beta}
    \beta(u) = e^{\theta_0 + \theta_x u_x + \theta_y u_y}, 
\end{equation}
where $(u_x, u_y)$ are the spatial coordinates of the location $u$, and $\theta_0$, $\theta_x$ and $\theta_y$ are the model parameters to be estimated. To illustrate the process, we compute the parameters for the ENS image shown in Figure~\ref{fig:human_ens}F. The hardcore distance of the ENS ganglia centers is $H$=3.6. In the fitted model, for interaction radius $R$=140, we obtain $\theta_0$=-8.19, $\theta_x$ = -7.52$\times$10\textsuperscript{-5}, $\theta_y$ = 2.57$\times$10\textsuperscript{-5} and interaction parameter $\gamma$=0.71. For generating random realizations of spatial point patterns from the fitted hardcore-Strauss model, we use the function \textit{spatstat::rmh.ppm}, which is based on the Metropolis-Hastings algorithm~\citep{geyer1994simulation, baddeley2000practical}. Figure~\ref{fig:simulated_ens}A shows a simulated realization of ENS ganglia centers generated from the fitted model, with the same sample window as the real human ENS ganglia centers (Figure~\ref{fig:human_ens}F). The L-function plot of the simulated realization is shown in Figure~\ref{fig:simulated_ens}B and it exhibits a spatial trend similar to the real ENS (compare with Figure~\ref{fig:human_ens}H).

We simulate the ENS network's realization by constructing the planar spatial network on top of the simulated ganglia centers. The vertices of this initial Delaunay triangulation have higher degrees than the real ENS networks (see Figure~\ref{fig:simulated_ens}C). Then, we sample edges from the triangulation using Algorithm~\ref{algo:1}. We experimented with three constraints imposed on sampling: the limitation in angles of the edges, the restriction in the length of the edges, use of both of them simultaneously. The evaluation showed that the simultaneous use of angle and length leads to realizations that are most similar (in the EMD metric) to those observed in real networks. Figure~\ref{fig:simulated_ens}D shows one example of the simulated ENS realization. Though there are a few high-degree vertices (with degrees 6 and 7), the vertices have an overall low degree (a mode of 3), similar to the real ENS network. All the degree-one vertices are located at the sample window boundary, recreating the sample dissection artifact. Figures~\ref{fig:simulated_ens}E-F show two more simulated realizations of the ENS network built on the same simulated ganglia centers. The distribution of the angle and the length of the edges of the simulated spatial networks are shown in Figures~\ref{fig:simulated_ens}G-H, respectively, along with the distributions of the real ENS network and the Delaunay triangulation-based network. The distribution of the edge angles and the edge lengths of the three simulated spatial networks are similar, and they are very close to the corresponding distributions of the real ENS network. The plots also illustrate that the starting planar network (Delaunay triangulation) is significantly different from the real one. As the rejection sampling process terminates, the angle and length distributions get closer to the real ENS network. 

\begin{figure}[t]
    \centering
    \includegraphics[width=\textwidth]{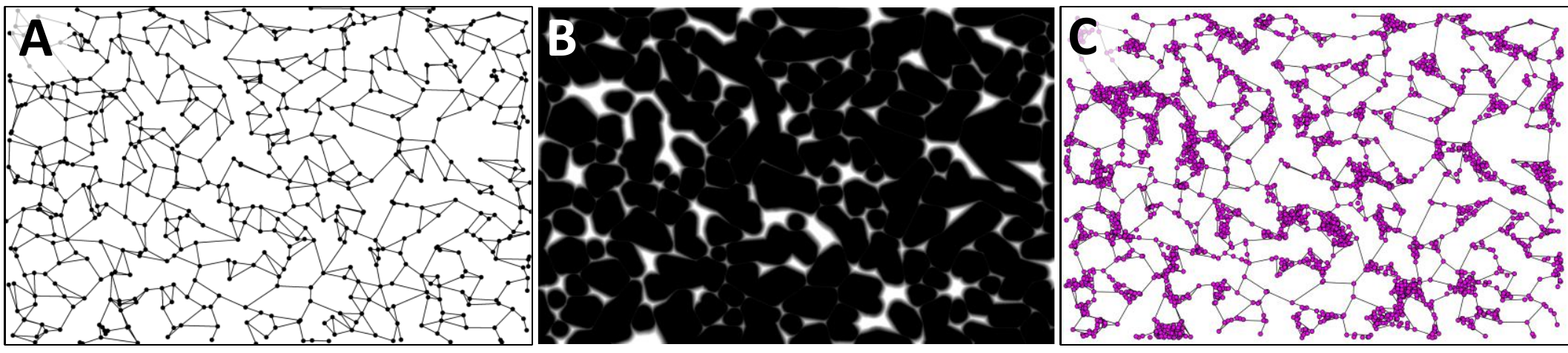}
    \caption{Experimental results for simulated realization of ganglionic neurons. (A) A simulated realization of the ENS network. (B) The greyscale intensity profile generated from (A). (C) A complete realization of the ENS network. The neurons (purple) are generated from a stationary HSP model and with subsequent thinning using the intensity profile in B.}
    \label{fig:intensity_profile_and_neuron}
\end{figure}

\begin{wrapfigure}[15]{r}{0.5\textwidth}
    \centering
    \includegraphics[width=0.5\textwidth]{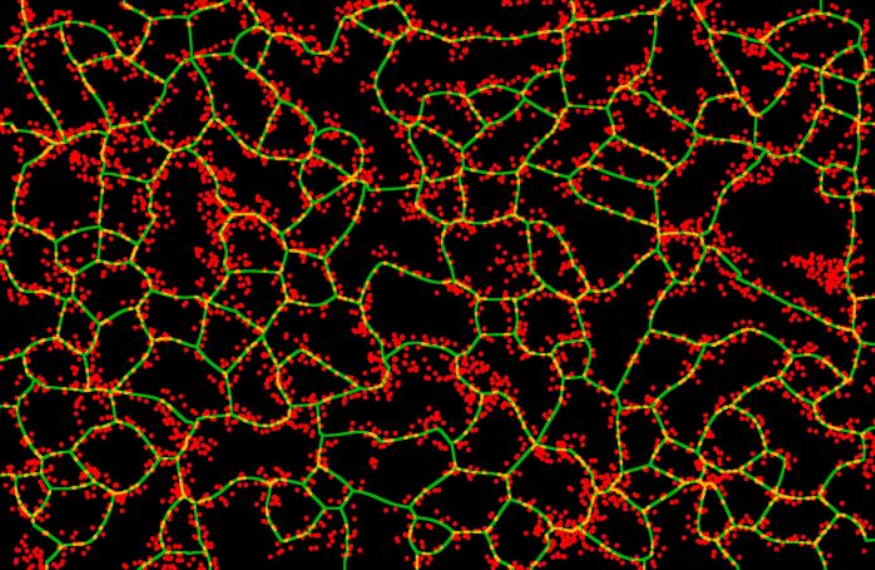}
    \caption{An illustration of a simulated ENS network, showing the neurons (red) and the inter-ganglionic connections (green).
    }
    \label{fig:abstract_neurons}
\end{wrapfigure}

To generate realizations of the neurons' centers, we construct the greyscale intensity profile using the simulated ENS network and generate a stationary HSP point pattern over the simulated ENS network's window. The parameter values are tuned by grid search. For the purpose of practical demonstration, we used the intensity parameter $\beta$=0.003 points per unit squared, interaction parameter $\gamma$=0.78, interaction radius $R$ = 0.0285 and hardcore distance $H$=0.01. Choosing different values for the HSP model parameters results in different spatial arrangements of the neurons, hence in different appearances of the ganglia. Finally, we introduce inhomogeneity by thinning the points with the greyscale intensity profile. Figures~\ref{fig:intensity_profile_and_neuron}A-C show a simulated realization of the ENS network, the generated intensity profile, and a complete ENS network with neurons, respectively. The correlation between the network structure and the arrangement of the simulated neurons is better visualized in Figure~\ref{fig:abstract_neurons}.
 
\subsection{Expressiveness of the generative model}

Table ~\ref{tab:ens_sample_data} shows the physiological information about the used set of human ENS samples. These samples are highly diverse in terms of age and health status. The table shows also the values of the selected parameters of our generative model: $\theta$ intensity coefficients, interaction parameter $\gamma$, interaction radius $R$, hardcore distance $H$ (see Equation~\ref{eq:cond_intensity} and Equation~\ref{eq:beta}) and average edge length of the network. The parameters $H$ and $R$ determine a disk-like neighborhood for each ganglia center. A ganglion mostly interacts (has inter-ganglionic connections) with other ganglia in its neighborhood, which means the average length of the inter-ganglionic connections should be close to $R$. This interpretation is reflected in the parameter values recorded in Table~\ref{tab:ens_sample_data}. We note some correspondence between the parameter values and the physiology. For instance, the interaction radius ($R$) is higher in adults than in the pediatric sample 2475-P. However, at this stage of research, it is difficult to discern whether the different ENS architectural parameters in different samples reflect distinct disease processes or age differences, or both. The generative model developed with sufficient ENS samples could give a range for interaction distance $(R)$ and intensity parameters ($\beta$ or $\theta$) that are separable in healthy adults and healthy children or healthy and diseased individuals. Due to the limited amount of biological data and difference in sample size, we cannot yet conclude that the recorded parameter values are the best-fitted ones for all ENS samples belonging to the same age group or having similar pathology. However, we argue that our generative model is expressive enough to capture the diversities in such ENS samples.   

\renewcommand{\arraystretch}{1.5}
\begin{table}[t]
    \centering \tiny
    \caption{Physiological condition of human patients and value of five selected parameters of the generative model of the ENS samples collected from their left colon. Sample status N: Normal, A: Affected. Average edge length, R and H are in \textmu m (see Table~\ref{tab:pixel-micron} for the pixel size of the sample ENS images). *$\gamma$=N/A and R=N/A indicates hardcore process with hardcore distance H.}
    \begin{tabular}{cccrrrcccc} 
    	\hline
    	\multirow{2}{*}{\textbf{Sample}} & \multirow{2}{*}{\textbf{Age}} & \multirow{2}{*}{\textbf{Primary condition}} & \multicolumn{3}{c}{\textbf{Intensity coefficients}}                                   & \multirow{2}{*}{$\gamma$} & \multirow{2}{*}{$\textbf{R}$} & \multirow{2}{*}{$\textbf{H}$} & \multirow{2}{*}{\textbf{Avg edge}}  \\
    	&                               &                                             & $\theta_0$ & $\theta_x$                          & $\theta_y$                         &                           &                               &                               &                                     \\ 
    	\hline
    	4443                             & 60 yo                         & Diverticulitis (N)                          & -11.70     & 1.35$\times$10\textsuperscript{-3}  & 8.37$\times$10\textsuperscript{-4} & 0.77                      & 378.8                         & 7.14                          & 304.65                              \\
    	4445                             & 28 yo                         & Diverticulitis (N)                          & -12.10     & 1.17$\times$10\textsuperscript{-3}  & 5.04$\times$10\textsuperscript{-4}                & 0.95                      & 397.71                        & 1.10                          & 309.46                              \\
    	4454                             & 36 yo                         & Diverticulitis (N)                          & -9.99      & -1.17$\times$10\textsuperscript{-3} & 7.05$\times$10\textsuperscript{-4}                 & 0.46                      & 231.68                        & 8.71                          & 204.40                              \\
    	4557                             & 37 yo                         & Sessile Polyp (N)                           & -14.20     & 6.28$\times$10\textsuperscript{-4}  & -2.10$\times$10\textsuperscript{-3}                & 0.82                      & 396.72                        & 4.34                          & 302.52                              \\
    	2475-P                           & 5~mo                          & Hirschsprung (A)                            & -9.10      & -4.74$\times$10\textsuperscript{-5} & 1.71$\times$10\textsuperscript{-5}                 & 0.71                      & 220.40                        & 5.67                          & 172.17                              \\
    	4598-P                           & 4~mo                          & Hirschsprung (A)                            & -9.92      & 3.84$\times$10\textsuperscript{-4}  & 2.40$\times$10\textsuperscript{-4}                 & N/A*                      & N/A*                          & 2.22                          & 106.24                              \\
    	\hline
    \end{tabular}
    \label{tab:ens_sample_data} 
\end{table}

\section{Conclusions and future work} 

We describe the initial development of a generative model of the enteric nervous system network. The model combines concepts from \textit{spatial point pattern analysis} and \textit{graph analysis}, two fields that have rarely been used together to solve practical problems from biological domains. In our model, we characterize the spatial location of ganglia and the neurons present in them with hybrid hardcore-Strauss processes (HSP). We describe the inter-ganglionic connections as spatial planar graphs and use measures such as the degrees of the vertices and angles and lengths of the edges to provide the spatial network characteristics. The model operates by generating realizations of ENS ganglia from the HSP fitted to actual biological samples. The edges are created by combining deterministic and random connection models. Finally, the model operation concludes by generating the location of ganglionic neurons using a second HSP. The computed features of the simulated spatial networks are similar to that of the real ENS networks. Our research demonstrated that the combined graph/SPP modeling technique presented, utilizing existing statistical approaches, offers a  parametric space suitable for describing a wide variety of real ENS organizations. Our models were appropriate for ENS architectures present in samples originating from adults, children, newborns, and healthy and sick individuals. The ability to represent ENS with a generative model dependent on a relatively low number of ante-hoc explainable parameters (some related to the SPP aspect and others from the connection model) will allow us to create compact and easy-to-process ENS feature vectors. We envision using our pipeline as a tool to rapidly and robustly characterize ENS architecture to define normal parameters and to identify disorganized ENS such as may occur in people with Hirschsprung disease, chronic intestinal pseudo-obstruction, or other severe bowel motility disorders. The proposed generative modeling approach and feature extraction from the model could complement traditional techniques used in pathology, which are based on qualitative descriptions, basic neuron counting, and other simple image analysis methods. These latter techniques might be operator-dependent and are often challenging to reproduce \citep{Swaminathan2010}. We see our modeling approach as a convenient alternative, which could be easily paired with computer-aided diagnostic (CAD) tools in downstream analysis. 

We are also aware of the model's limitations. Firstly, the number of accessible clinical samples utilized in the model development and validation is relatively low. This is due to the difficulty in obtaining the biological samples and technological challenges associated with the pre-processing required for sophisticated laser-scanning microscopy imaging. Consequently, we face a tough choice of whether the model development should be focused on a narrow range of conditions and patients or should be aimed at a more generalized ENS description. The first route would allow a demonstration of applicability in computer-aided diagnostics, but it would also be very self-limiting. Secondly, our model does not capture the diversity of neurons and glial cells present in ENS in the current version. It focuses on the architectural rather than functional properties of the network. Therefore, any biological conclusions based on the ENS descriptions recovered by the model must be paired with cell-functional descriptions obtained using immunostaining and manual examination of the samples conducted by neuroscientists and pathologists. Naturally, we plan to expand our model to multiple cell types.

Despite these limitations, we believe that our approach opens the door to a significant breakthrough in the use of ENS imaging for diagnostics. By introducing a well-defined procedure for characterizing  ENS networks to describe overall network properties, including the neuronal clusters of ganglia and the inter-ganglionic connections, we are moving ahead from observer-dependent scoring and qualitative descriptors of ENS complexity, which do not readily capture network architectural features defined in our models.

\section*{Conflict of Interest Statement}
%All financial, commercial or other relationships that might be perceived by the academic community as representing a potential conflict of interest must be disclosed. If no such relationship exists, authors will be asked to confirm the following statement: 
The authors declare no competing interests.

\section*{Author Contributions}
BR conceived and planned the study; AP contributed to the mathematical models; MJH, JDE, ROH collected and processed the biological samples. ASS and BR executed the study and co-wrote the manuscript with input from all the researchers.

\section*{Funding}
This work was supported by NIH SPARC (Stimulating Peripheral Activity to Relieve Conditions) Programs OT2TR001965 (Bartek Rajwa and Abida Sanjana Shemonti), and OT2OD023859 (Marthe J. Howard and Robert O. Heuckeroth). Robert O. Heuckeroth is also supported by the Irma and Norman Braman Endowed Chair for Research in GI Motility Disorders and by the Suzi and Scott Lustgarten Center endowment.

%\section*{Acknowledgments}

\section*{Supplemental Data}

\begin{table}[h]
    \centering 
    \caption{The pixel size of the sample ENS images in \textmu m.}
    \begin{tabular}{lr}   
    \\ \hline 
        Sample Id & Pixel size (\textmu m)\\
    \hline 
        4443 & 0.947\\
        4445 & 0.8303\\
        4454 & 0.7085\\
        4557 & 1.5743\\
        2475-P & 1.5743\\
        4598-P & 1.5743\\
    \hline  
    \end{tabular} 
    \label{tab:pixel-micron} 
\end{table}

%\section*{Data Availability Statement}
% Please see the availability of data guidelines for more information, at https://www.frontiersin.org/about/author-guidelines#AvailabilityofData

\bibliographystyle{Frontiers-Harvard} %  Many Frontiers journals use the Harvard referencing system (Author-date), to find the style and resources for the journal you are submitting to: https://zendesk.frontiersin.org/hc/en-us/articles/360017860337-Frontiers-Reference-Styles-by-Journal. For Humanities and Social Sciences articles please include page numbers in the in-text citations 
\bibliography{ref-list}

%%% Make sure to upload the bib file along with the tex file and PDF
%%% Please see the test.bib file for some examples of references

%%% Please be aware that for original research articles we only permit a combined number of 15 figures and tables, one figure with multiple subfigures will count as only one figure.
%%% Use this if adding the figures directly in the mansucript, if so, please remember to also upload the files when submitting your article

\end{document}